\documentclass[trackchanges,twocolumn]{aastex701}

\usepackage{gensymb}
\usepackage{rotating} 
\usepackage{amsmath}
\usepackage{comment}
\usepackage{gensymb}

\usepackage{graphicx}
\usepackage{subcaption}

\begin{document}

\title{Evidence for an Atmosphere on the Ultra-Short Period super-Earth HD~3167~b}

\author[0000-0002-0508-857X]{Brandon Park Coy}
\affiliation{Department of the Geophysical Sciences, University of Chicago, Chicago, IL, USA}
\email[show]{bpcoy@uchicago.edu}  

\author[0000-0002-6215-5425]{Qiao Xue}
\affiliation{Department of Astronomy \& Astrophysics, University of Chicago, Chicago, IL, USA}
\email{qiaox@uchicago.edu}

\author[0000-0003-4241-7413]{Megan Weiner Mansfield}
\affiliation{Department of Astronomy, University of Maryland, College Park, MD, USA}
\email{mwm@umd.edu}

\author[0000-0003-3773-5142]{Jason D.\ Eastman}
\affiliation{Center for Astrophysics \textbar \ Harvard \& Smithsonian, Cambridge, MA, USA}
\email{jason.eastman@cfa.harvard.edu}

\author[0000-0002-4487-5533]{Anjali A. A. Piette}
\affiliation{School of Physics and Astronomy, University of Birmingham, Birmingham, UK}
\email{a.a.a.piette@bham.ac.uk}

\author[0000-0002-0692-7822]{Tyler Fairnington}
\affiliation{Department of Astronomy \& Astrophysics, University of Chicago, Chicago, IL, USA}
\email{tfairnington@uchicago.edu}

\author[0009-0004-1094-3093]{Cole Smith}
\affiliation{Department of Astronomy, University of Maryland, College Park, MD, USA}
\email{csmit103@umd.edu}

\author[0000-0002-0659-1783]{Michael Zhang}
\affiliation{Department of Astronomy \& Astrophysics, University of Chicago, Chicago, IL, USA}
\email{mzzhang2014@gmail.com}

\author[0000-0002-1337-9051]{Eliza M.-R. Kempton}
\affiliation{Department of Astronomy \& Astrophysics, University of Chicago, Chicago, IL, USA}
\email{ekempton@uchicago.edu}

\author[0000-0003-4733-6532]{Jacob L.\ Bean}
\affiliation{Department of Astronomy \& Astrophysics, University of Chicago, Chicago, IL, USA}
\email{jacobbean@uchicago.edu}

\author[0000-0002-1592-7832]{Xuan Ji}
\affiliation{Department of the Geophysical Sciences, University of Chicago, Chicago, IL, USA}
\email{xuanji@uchicago.edu}

\author[0000-0002-8518-9601]{Peter Gao}
\affiliation{Earth and Planets Laboratory, Carnegie Institution for Science, Washington DC, USA}
\email{pgao@carnegiescience.edu}

\author[0000-0003-2775-653X]{Jegug Ih}
\affiliation{Space Telescope Science Institute, Baltimore, MD, USA}
\email{jih@stsci.edu}

\author[0000-0002-9076-6901]{Daniel D.B. Koll}
\affiliation{Department of Atmospheric and Oceanic Sciences, Peking University, Beijing, People's Republic of China}
\email{dkoll@pku.edu.cn}

\author[0000-0002-4671-2957]{Rafael Luque}
\affiliation{Instituto de Astrof\'isica de Andaluc\'ia (IAA-CSIC), Glorieta de la Astronom\'ia s/n, Granada, Spain}
\email{rluque@iaa.es}

\author[0000-0003-2066-8959]{Jaume Orell-Miquel}
\affiliation{Department of Astronomy, University of Texas at Austin, Austin, TX, USA}
\email{jaume.miquel@austin.utexas.edu}


\author[orcid=0000-0002-1426-1186]{Edwin S. Kite}
\affiliation{Department of the Geophysical Sciences, University of Chicago, Chicago, IL, USA}
\email{kite@uchicago.edu}

\begin{abstract}
`Lava worlds'--Earth-sized planets hot enough ($T_{eq}\gtrsim1100$\,K) to melt their dayside silicate surfaces--have emerged as promising candidates for atmospheric detection and characterization. Thermal emission observations show an apparent dichotomy: the hottest lava worlds have colder daysides than the temperature of a maximally emitting bare rock, indicating the likely presence of thick and/or reflective atmospheres while the coldest ones do not. However, where in instellation flux this potential bifurcation occurs is uncertain. 
We present a JWST MIRI LRS eclipse of the ultra-short period (USP) lava world HD~3167~b ($T_{eq}=1786$\,K, $R=1.6\,R_{\oplus}$, $P=0.96$\,d) that helps bridge this gap. 
We measure the white light eclipse depth to be $38\pm11$\,ppm, more than 5\,$\sigma$ lower than the expected eclipse depth of a dark, maximally hot bare rock. We use this to derive a dayside brightness temperature that is best explained by the presence of an atmosphere that cools the dayside by reflecting incoming starlight and/or efficiently redistributing heat to the planet's nightside. An atmosphere is further compatible with the planet's slight under-density compared to an Earth-like composition. The corresponding dayside emission spectrum is not precise enough to constrain atmospheric composition, motivating follow-up spectroscopic observations with JWST NIRSpec. 
Lastly, we use our observation and existing data to refine key planetary parameters of the HD~3167 system.  HD~3167~b is currently the least irradiated USP super-Earth with evidence for an atmosphere. 

\end{abstract}



\section{Introduction} \label{sec:intro} 

Uncovering the prevalence of and conditions required for atmospheres on terrestrial planets is a key goal of astrophysics.  While most thermal emission observations of Earth-sized planets around small M stars indicate bare rock surfaces (e.g., \citealt{coy2025population,kreidberg2025first}), \textit{lava worlds}, small ($\lesssim1.9\,R_{\oplus}$) planets hot enough to melt their dayside surfaces, have emerged as promising candidates for atmospheric characterization. Here, we define `lava world' as a small planet with a substellar temperature hot enough to fully melt rock of Earth-like composition (roughly $\sim1500\,$K, \citealt{lutgens00}), although this exact threshold depends on surface composition and thus is a loose definition.  Lava worlds' extremely short tidal circularization and tidal locking timescales compared to their generally advanced ages \citep{winn2018kepler} make them prime candidates for using measurements of the dayside temperature via secondary eclipse as a probe for atmosphere presence or absence \citep{koll2019identifying,mansfield2019identifying}.  Four lava worlds thus far --- TOI-431 b \citep{monaghan2025low}, 55 Cancri e \citep{hu2024secondary}, K2-141 b \citep{zieba2022k2}, and TOI-561 b \citep{teske2025thick} --- show dayside brightness temperatures much lower than expected for an airless body, suggesting efficient heat transport to the nightside, molecular absorption features from gases, and/or high Bond albedo cooling their daysides, all signatures of atmospheres. Explanations of these observations that do not involve atmospheres are unlikely, as partially or fully molten silicates are expected to have low reflectivity \citep{essack2020low}, and the heat transport from magma currents is too weak to significantly cool the dayside \citep{kite2016atmosphere}. These detections challenge the `Cosmic Shoreline' concept \citep{zahnle2017cosmic}, which predicts that these ultra-irradiated planets cannot retain significant atmospheres due to their highly elevated cumulative atmospheric loss fluxes.  While theory predicts thin persistent atmospheres on lava worlds formed via vapor pressure equilibrium with the underlying magma pool (e.g., \citealt{kite2016atmosphere,curry2025chemical}), the heat redistribution expected from these thin atmospheres is likely too small to explain current observations.

The current data point to a bifurcation in lava worlds with and without atmospheres, but exactly where this potential transition from `bare rock' surface to a thick and/or reflective atmosphere occurs, and why, is uncertain.  Previous observations have suggested a potential `transition irradiation temperature' ($T_{irr}$, the sub-stellar temperature of a perfect blackbody) between 1950\,K and 2600\,K; GJ 367 b, a dense sub-Earth around an M star with $T_{irr}$\,=\,1950\,K seemingly lacks an atmosphere \citep{zhang2024gj}, whereas observations of TOI-431 b, an Earth-sized planet with $T_{irr}$\,=\,2600\,K around a K star are inconsistent with a dark, bare rock, suggesting the presence of an atmosphere \citep{monaghan2025low}.

HD~3167~b \citep{vanderburg2016two} is an ultra-short period (USP), $1.6\,R_{\oplus}$ super-Earth lava world with $T_{irr}\sim2500\,$K that can help bridge this gap.  Its bulk density ($\rho=5.5\pm0.5$\,g/cm$^3$, \citealt{bourrier2022cheops}) is lower than expected for an Earth-like composition ($\sim8$\,g/cm$^{3}$ for a $1.6\,R_{\oplus}$ planet including compression effects, \citealt{zeng2019}), suggesting either the lack of an iron core or a thick, extended atmosphere.  This degeneracy can be investigated through measurements of the planet's secondary eclipse depth, which can in turn constrain the amount of heat redistribution or Bond albedo of an atmosphere or surface \citep{koll2019identifying,mansfield2019identifying}. HD~3167~b's high emission spectroscopy metric (ESM, \citealt{kempton2018framework}) of 13 offers an excellent opportunity to probe the atmospheric properties of a low-density USP super-Earth.

In this Letter, we present a JWST MIRI LRS observation of HD~3167~b's secondary eclipse. We describe the observation and data analysis in Section \ref{sec:observations}, our interpretation in Section \ref{sec:interp}, and present our conclusions and room for future work in Section \ref{sec:conc}.

\section{Observation and Data Analysis} \label{sec:observations}
We observed a single MIRI LRS eclipse of HD~3167~b on June 25th, 2025 as part of JWST GO Program 4818 (PI: Megan Weiner Mansfield), also known as LAVA LAMPS (Looking At Vaporized Atmospheres of LAva/Magma Planets Survey). The observation consisted of a 4.3-hr exposure and a 10-integration exposure offset from the target aimed at characterizing the contemporaneous mid-infrared background.  These exposures used the FASTR readout mode with 9 groups per integration (1.6\,s per integration) for a total of 9646 integrations.  The observations were designed to not surpass 65\% saturation (50\% of full well).The observation was scheduled assuming the eclipse occurs at 0.5 orbital phase, and previous eccentricity constraints give an uncertainty in eclipse timing of only $\sim0.014$ days, or 20 minutes \citep{bourrier2022cheops}.  The host star HD~3167 is a bright ($\mathbf{m_{K_{s}}}=7.07$), old ($\sim10$\,Gyr, \citealt{bourrier2022cheops}) K0 V star considered to be chromospherically inactive \citep{vanderburg2016two}, making it an optimal target for secondary eclipse observations with JWST. To ensure robustness and reproducibility of our results, we use three widely-used independent data reduction pipelines, \texttt{SPARTA}, \texttt{Eureka!}, and \texttt{exoTEDRF} to analyze the data. 

\subsection{\texttt{SPARTA}}

\texttt{SPARTA} \citep{kempton2023reflective} is an end-to-end data reduction pipeline completely independent of the standard \texttt{jwst} pipeline, and has been used in several previous analyses of MIRI LRS data \citep[e.g., ][]{xue2024jwst,mansfield2024no,zhang2024gj}.  Our analysis follows that of \citet{mansfield2024no} with some modifications.

We begin with the uncalibrated data by performing nonlinearity correction, dark subtraction, multiplication by the gain (fixed to 3.1 electrons per data number, \citealt{kempton2023reflective}), two rounds of up-the-ramp fitting to address MIRI LRS nonlinearities, and flat fielding.  We tested using our dedicated background exposure versus the standard flat field image and saw no noticeable difference in the resulting light curve. We discard the last group known to be anomalous due to the `last frame effect' of MIRI (e.g., \citealt{morrison2023jwst}).  We tested including a version of the \texttt{emicorr} step (M. Weiner Mansfield et al., in prep.) from the \texttt{jwst} pipeline to mitigate the $\sim$390~Hz and $\sim$10~Hz periodic electronic noise patterns seen in MIRI LRS \citep{bell2024nightside}.  However, we noticed only a negligible improvement in the median absolute deviation (MAD) of residuals in the resulting light curve (0.3\%) and very similar retrieved parameters and thus forgo this step in our reduction.  \texttt{SPARTA} rotates the MIRI LRS image so that the wavelength axis spans the $x$ direction.  We then remove the background at each wavelength column by subtracting the median of rows 10:25 and -25:-10 for each column. We then calculate a median image across all integrations and use it as a template to calculate the relative per-integration position offset of the spectral trace in the $x$ and $y$ directions. Using the previously calculated median image to generate a spatial profile, we use optimal extraction \citep{horne1986optimal} to extract a spectrum for each integration. We use an extraction half-width of 5 pixels and reject pixels $>$5$\,\sigma$ away from the median image as outliers.  We tested different extraction half-widths varying from 5 to 10 pixels and found that it had little effect on retrieved white light curve parameters. The optimally-extracted spectra for each integration are gathered into one pickle file. Integrations that are $>$4$\,\sigma$ outliers in the white light curve are removed, while $>$4$\,\sigma$ outliers in the unbinned spectroscopic light curves are repaired via 2-D interpolation.

\subsection{\texttt{Eureka!}}
\texttt{Eureka!}\ \citep{bell_eureka_2022} uses the official \texttt{jwst} calibration pipeline and processes raw, uncalibrated \texttt{FITS} files through six modular stages. We used \texttt{Eureka!}\ version 1.3 with CRDS context 1477. Below, we describe our implementation that deviates from the standard sequence.

In stage~1, we included the \texttt{emicorr} step within the \texttt{jwst} pipeline--including this step results in a 12\% reduction in the MAD of the residuals in the resulting light curve.  We also enabled the group-level background subtraction step within \texttt{Eureka!}, which subtracts a per-group background prior to ramp fitting. Background regions were defined along the spatial axis, spanning pixels~0--17 on one side of the trace and pixels~55--71 on the other, leaving a 37-pixel space that safely brackets the star's point spread function (PSF). 

In stage~3, the data were trimmed to the subarray region $y=[60, 393]$ and $x=[11, 61]$.  The centroid of the spectral trace was determined for each integration by fitting a Gaussian profile along the spatial direction.
For each integration, a constant was fit to the background region with individual pixels deviating by more than $5\sigma$ from the model masked. The resulting background model was subtracted from the full 2D spectral image. We then performed optimal spectral extraction \citep{horne1986optimal}. The spatial profile was constructed from the median stack of all integrations, smoothed with a window of 20~pixels. Outliers in the median frame that deviated by $>$10$\,\sigma$ were masked. To select the optimal extraction and background aperture widths, we performed a grid search over extraction half-widths of 4, 5, 6, 7, and 8~pixels and background
exclusion half-widths of 12, 13, 14, and 15~pixels, evaluating each combination by the MAD of the resulting white light curve residuals. The combination of an extraction half-width of 5~pixels and a background half-width of 13~pixels minimized the MAD and was adopted for all subsequent reductions. 

In stage~4, the time-series of spectra is binned in wavelength to produce light curves. We manually masked column 369 on the 2-D light curve as it showed abnormal behavior. In each light curve, data points that deviate by more than 5$\,\sigma$ from a rolling median were discarded.

\subsection{\texttt{exoTEDRF}}
We also performed an independent reduction using the \texttt{exoTEDRF} (formerly \texttt{supreme-SPOON}) pipeline \citep{radica2024exoTEDRF}, whose capabilities have recently been extended to MIRI LRS data \citep{luque2025insufficient}. \texttt{exoTEDRF} performs the reduction in three stages: detector-level processing, spectroscopic processing, and 1D spectral extraction. Stage 1 largely serves as a wrapper for the standard \texttt{jwst} pipeline, with minor corrections tailored to the MIRI LRS systematics. In particular, we apply the \texttt{emicorr} step to mitigate the $\sim$390 Hz row-correlated noise \citep{bell2024nightside}, and correct for the reset switch charge decay nonlinearity \citep{morrison2023jwst} by flagging the first 12 up-the-ramp groups. Cosmic ray jumps are corrected in the \texttt{JumpStep} and are identified through time-domain outlier rejection, with $>$5$\,\sigma$ clipping used to mask affected pixels.  

Background subtraction in stage 2 is performed at the integration level rather than the group level.  We estimate the row-by-row background using the median in columns 12 -- 26 and 46 -- 60, masking the spectral trace with a 14-pixel-wide aperture and using 10 columns on either side to define the background regions. We correct bad pixels in both the spatial and temporal domains. For the spatial correction, we identify $>$5$\,\sigma$ outliers in the median-stacked image and replace them using the median of neighboring pixels. An analogous interpolation is then applied in the temporal direction, flagging $>$5$\,\sigma$ outliers, and using the same method as the \texttt{JumpStep} step. In the final spectroscopic calibration, we perform a Principal Component Analysis (PCA) reconstruction of the 3D flux cube (integrations, x-pixel, y-pixel, see \citealt{luque2025insufficient}) to remove components that indicate detector trends. We remove the third component, which correlates with a horizontal drift of the spectral trace over time, resulting in a 12\% decrease in MAD of the white light curve residuals compared to not including this step.

In stage 3, we extract the 1D time-series spectra using box aperture extraction, where the location of the spectral trace is determined with the \texttt{edgetrigger} \citep{Radica:2022} algorithm. To optimize the extraction aperture, we performed a parameter sweep over box widths from 3 to 12 pixels and selected a final extraction width of 4 pixels, which yielded the highest white light curve precision.

\subsection{Light Curve Fitting}

For our white light analyses, we sum the data over 5.06--10.55 $\mu$m, motivated by the dropoff in MIRI LRS throughput and increase in thermal background at longer wavelengths, and the mismatch between the observed absolute flux and stellar model fluxes at wavelengths shorter than 5.06\,$\mu$m (e.g., \citealt{Zieba2026}).  We also notice a sharp shift in the systematic ramp behavior (downward to upward) in the `shadowed region' (10.55--11.77\,$\mu$m, \citealt{bell2024nightside,zhang2024gj}) that would bias the white light results.

\begin{figure*}
    \centering
    \includegraphics[width=0.8\linewidth]{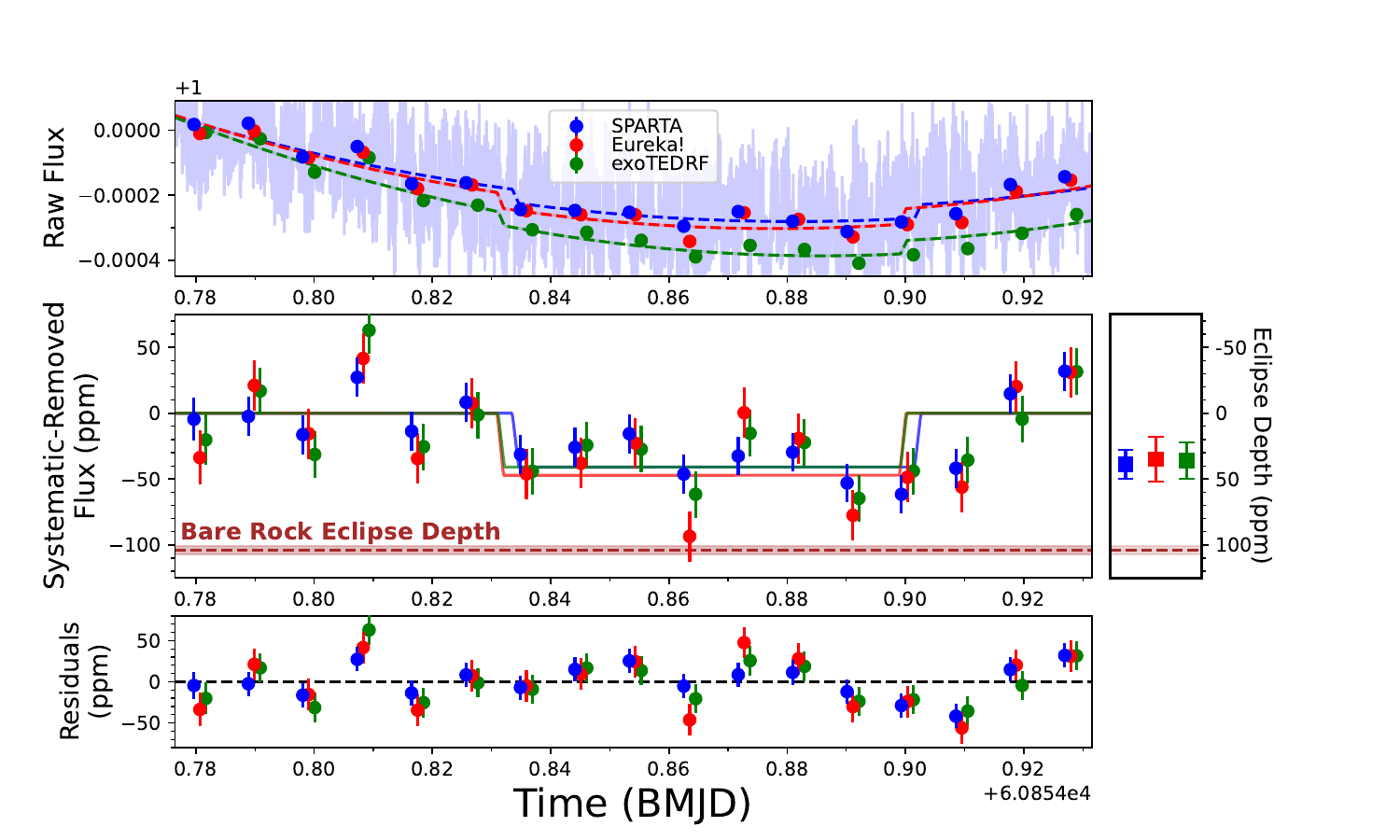}
    \caption{Best-fit white light (5.06--10.55\,$\mu$m) curves of HD~3167~b's secondary eclipse from our \texttt{SPARTA} (blue), \texttt{Eureka!} (red), and \texttt{exoTEDRF} (green) reductions. (Top) Binned raw white light fluxes.  The best-fit full \texttt{SPARTA} model (with integration-level $x$ decorrelation), \texttt{Eureka}, and \texttt{exoTEDRF} models are shown alongside the quadratic ramp component shown as dashed lines. (Middle) Binned fluxes after dividing by the best-fit systematics model for each reduction.  Retrieved eclipse depths and uncertainties are shown on the right. (Bottom) Binned residuals of the best-fit full models.  For all plots, \texttt{Eureka!}/\texttt{exoTEDRF} data are offset by 0.001/0.002 days for visual clarity.}
    \label{fig:lightcurves}
\end{figure*}

To model long-term trends in the white light curve (we discuss this choice in Appendix Section \ref{ap:sys}), we use the following quadratic ramp model,
\begin{equation}
    F_{sys}=F_{star}\times(1+A(t-t_{min})^2+B(t-t_{min})+c_{x}x),
\end{equation}
where $t_{min}$ represents the beginning time of the trimmed light curve, $F_{star}$ is a normalization constant, and $c_{x}$ decorrelates with the $x$ position of the spectral trace (referred to as $y$ in \texttt{Eureka!}). We do not find a significant correlation with the $y$ position or $x$/$y$ widths of the spectral trace and thus do not include them in our fits.  For \texttt{Eureka!} and \texttt{exoTEDRF}, we do not include decorrelation with $x$ as 1) we do not see statistically significant correlation in \texttt{Eureka!} and 2) our PCA reconstruction with \texttt{exoTEDRF} roughly models out correlation with the trace position.  All fitting is done at native time resolution (1.6 s per integration) with no binning.  We trim the first 1200 integrations (32 minutes) when fitting each light curve, determined by optimizing goodness-of-fit and information criteria (see Appendix Section \ref{ap:sys}).
Our astrophysical model is determined using \texttt{batman} 
\citep{kreidberg2015batman}.  We tested including a phase modulation term due to the planet's short orbital period, where
\begin{equation}
    F_{p}(\theta)=F_{p,batman}\times\left(\frac{1}{2}+\frac{1}{2}\cos{\theta}\right),
\end{equation}
where $\theta$ is the planet's orbital phase (0 at mid-eclipse), but note that the planet's signal is only expected to vary by $\lesssim$12\% over our observation, and we retrieve identical eclipse depths with or without this phase modulation term.  This is likely due to the planet's real phase modulation being absorbed into the systematics model.  We include modifications to \texttt{SPARTA} to model the light-time travel delay, roughly $2a/c\sim$18 seconds for our observation, by fixing the host star radius at $R=0.865\,R_{\odot}$ \citep{bourrier2022cheops}. 
We tested allowing a free eccentricity for these fits but ultimately decided that the signal-to-noise of the eclipse is too low (due to only being detected at $\sim2-3.5\,\sigma$) to accurately constrain a potential non-zero eccentricity (see Appendix Section \ref{ap:ecc}), and thus adopt a circular model ($e=0$) for our fiducial analyses.

For posterior sampling, we use the nested sampling code \texttt{dynesty} \citep{speagle2020dynesty} with an initial convergence criterion of $\Delta\log Z=0.1$. We use dynamic nested sampling (with 1000 initial walkers), as it is designed to better describe the true posterior distribution than static nested sampling. We fix orbital and planetary parameters to the median values presented in \citet{bourrier2022cheops} with a Gaussian prior on the mid-eclipse time.  Each fit includes an error inflation factor that inflates the pipeline-reported uncertainties to ensure that we are not overestimating our precision on the eclipse depth.
We note that including informative Gaussian priors on orbital parameters retrieves nearly the exact same eclipse depths (with only a $\sim1\%$ increase in uncertainty), as these parameters are measured very precisely \citep{bourrier2022cheops}. Our priors and posteriors used in light curve fitting are shown in Table \ref{tab:lightcurve}, with our best-fit white light curves shown in Figure \ref{fig:lightcurves}.  We reach an eclipse depth precision with \texttt{SPARTA} near the expected photon-limited precision based on JWST ETC output ($\sim11$\,ppm).

\begin{deluxetable*}{lccccc}
\tabletypesize{\scriptsize}
\tablenum{1}
\tablecaption{Priors and Posteriors from LRS White Light Curve Fitting}\label{tab:lightcurve}
\tablewidth{0pt}
\tablehead{
\colhead{Parameter} & \colhead{Prior} & \colhead{Posterior} & \colhead{Posterior} &  \colhead{Posterior} & \colhead{\textbf{Posterior (Adopted)}}\\
\colhead{} & \colhead{\texttt{SPARTA/Eureka!/exoTEDRF}} & \colhead{\texttt{SPARTA}} & \colhead{\texttt{Eureka!}} & \colhead{\texttt{exoTEDRF}} & \colhead{\texttt{ExoFASTv2} Global Fit}}
\startdata
    Period (d) & 0.95965428$^{a}$ & --  & -- & -- & $\mathbf{0.95965451^{+2.2e-7}_{-2.1e-7}}$ \\
    Semi-major Axis ($a/R_{\star}$) & 4.450$^{a}$  & --  & -- & -- & $\mathbf{4.525^{+0.069}_{-0.110}}$\\
    Inclination ($\degree$)  & 87.59$^{a}$ & --  & -- & -- & $\mathbf{87.8^{+1.5}_{-1.6}}$ \\
    Radius ($R_{p}/R_{\star}$) & 0.01712$^{a}$ & --   & -- & -- &$\mathbf{0.01698^{+0.00019}_{-0.00018}}$\\ 
    Mid-Eclipse Time (BMJD) & $\mathcal{N}(60854.8674,0.0011)$$^{a}$  &  $60854.8673^{+0.0008}_{-0.0009}$  & $60854.8661^{+0.0021}_{-0.0007}$ & $60854.8672^{+0.0006}_{-0.0014}$ &$\mathbf{60854.8677\pm0.0012}$\\   
    Eclipse Depth (ppm) & $\mathcal{U}(-1000,1000)$ &  $39\pm11$ & $35\pm17$  &  $36\pm14$ &$\mathbf{38\pm11}$ \\
\enddata
\tablecomments{$^{a}$Median values taken from \citet{bourrier2022cheops}. The mid-eclipse time prior is determined by propagating uncertainties on the orbital ephemeris assuming an orbital phase of 0.5.}
\end{deluxetable*}

\subsection{Global System Fit} \label{sec:global}

Precise planetary parameters are key to interpreting the eclipse depth and potential atmospheric composition of small planets.  The HD~3167 system has been studied by Kepler \citep{vanderburg2016two}, Hubble \citep{guilluy2020ares}, Spitzer IRAC Channel 2 (4.5\,$\mu$m) \citep{mikal2020transmission}, CHEOPS \citep{bourrier2022cheops}, and TESS.  The system has also been monitored over a 5.5 year timespan by ground-based radial velocity (RV) instruments, including Keck/HIRES, APF/Levy, ESO3.6m/HARPS, and TNG/HARPS-N.

We use \texttt{ExoFASTv2} \citep{Eastman:2019,mahajan2024using} to jointly-fit photometry from Kepler, Hubble, Spitzer, CHEOPS and RV data presented in \citet{bourrier2022cheops} and \citet{bonomo2023cold} to derive precise planetary parameters of the HD~3167 system. Detrended Kepler, Hubble, Spitzer, and CHEOPS light curves are taken from \citet{bourrier2022cheops} using their inflated errorbars (to account for uncertainties in detrending parameters).  RV data from \citet{bonomo2023cold} are normalized by fitting a median offset between overlapping observations with \citet{bourrier2022cheops}.  Following \citet{bourrier2022cheops}, we detrend the HARPS-N RV data linearly with the reported full-width half-maximum of the cross-correlation function. We also add photometry from TESS sector 70 using a spline function to detrend stellar variability and our JWST MIRI LRS white light curve. Contrary to \citet{bourrier2022cheops}, we fit for a single planet radius shared among different instruments/wavelength ranges for HD~3167~b and c (separately). HD~3167~b likely lacks a hydrogen-dominated atmosphere where absorption features would significantly affect its radii in different bandpasses, and HD~3167 c possess a very high metallicity  ($\gtrsim700\times$ solar, \citealt{mikal2020transmission}) atmosphere with a small scale height, showing good agreement in transit depth over multiple instruments.  Note that we do not use the empirical mass-luminosity-metallicity relation for low-mass stars derived in \citet{mann2019constrain} used in previous similar fits \citep{mansfield2024no,xue2024jwst} due to the Sun-like nature of the host star.  We instead directly joint-fit MIST isochrones \citep{Dotter2016,Choi2016}, the stellar spectral energy distribution (SED), Gaia parallax, and transit/eclipse-derived stellar density to derive stellar properties.

For this global fit, we use our \texttt{SPARTA} reduction of the eclipse data due to the notably higher precision achieved with \texttt{SPARTA} compared to \texttt{Eureka!} for multiple similar observations (e.g., \citealt{mansfield2024no,xue2024jwst}).  The MAD of the best-fit light curve is also the lowest of the three reductions by a significant margin (\texttt{SPARTA}: 221\,ppm, \texttt{Eureka!}: 289\,ppm, \texttt{exoTEDRF}: 267\,ppm).  We retrieve an eclipse depth of $38\pm11$ ppm, fully consistent with all three of our reductions. For reference, the blackbody LRS white light eclipse depth expected from our new global system fit is $104\pm3$ ppm. Our global fit posteriors are also reported in Table \ref{tab:lightcurve}, with full results in Appendix Section \ref{ap:exofast}.

\subsection{Spectral Light Curves}
We extracted spectral light curves by fitting an identical systematics model (per spectral bin) to our white light analyses.  Due to very low SNR, we use large 1.83 $\mu$m bins ([5.06,6.89], [6.89,8.72], [8.72,10.55]\,$\mu$m).  We include a separate 1.22 $\mu$m bin (10.55--11.77\,$\mu$m) to model the  `shadowed region' \citep{bell2024nightside,zhang2024gj} that often exhibits significant differences in systematics.  For these fits, we fix the orbital parameters ($P$, $i$, $a/R_{\star}$, $R_{p}/R_{*}$) to the best-fit parameters of our global \texttt{ExoFASTv2} fit, using our best-fit \texttt{ExoFASTv2} mid-eclipse time for \texttt{SPARTA} and \texttt{Eureka!}/\texttt{exoTEDRF} mid-eclipse time for \texttt{Eureka!}/\texttt{exoTEDRF}, respectively.  
Results are reported in Table \ref{tab:spectrum}. However, we caution against over-interpretation of these spectra as the uncertainties are large, depths are sensitive to the assumed mid-eclipse time, and spectra are discrepant between pipelines up to the $\sim2.5\,\sigma$ level.

\begin{deluxetable*}{lcccc}
\tabletypesize{\small}
\tablecaption{Low Resolution Emission Spectrum of HD~3167~b}
\tablenum{2}\label{tab:spectrum}
\tablewidth{0pt}
\tablehead{
\colhead{$\lambda_{start}$ ($\mu$m)} & \colhead{$\lambda_{end}$ ($\mu$m)} & \colhead{\texttt{SPARTA} Depth (ppm)} & \colhead{\texttt{Eureka!} Depth (ppm)} & \colhead{\texttt{exoTEDRF} Depth (ppm)}}
\startdata
     5.06 & 6.89 & $19\pm14$ & $48^{+18}_{-19}$ & $37\pm18$ \\
     6.89 & 8.72 & $72^{+19}_{-20}$ & $25\pm24$ & $52^{+24}_{-23}$\\
     8.72 & 10.55 & $30^{+29}_{-30}$ & $11^{+39}_{-38}$ & $38^{+34}_{-35}$ \\
     10.55 & 11.77 & $61^{+79}_{-80}$ &$79^{+106}_{-104}$ & $180\pm96$  \\
\enddata
\end{deluxetable*}

\section{Interpretation} \label{sec:interp}

\subsection{Brightness Temperature Ratio}

We follow the methods outlined in \citet{xue2024jwst} to fit for a `brightness temperature ratio' $\mathcal{R}$, defined as the measured dayside brightness temperature $T_{day}$ divided by that of a zero-albedo, zero-heat redistribution blackbody $T_{max}$, or 
\begin{equation}
\mathcal{R}\equiv\frac{T_{day}}{T_{max}}=\left(\frac{2}{3}\right)^{-1/4}(1-A_{eff})^{1/4}\left(\frac{2}{3}-\frac{5}{12}\varepsilon\right)^{1/4} ,
\end{equation}
where,

\begin{equation}
    T_{max}=T_{\star}\sqrt{\frac{R_{\star}}{a}}\left(\frac{2}{3}\right)^{1/4}.
\end{equation}
$\varepsilon$ is the heat redistribution efficiency, $A_{eff}$ is the effective albedo, $a$ is the planet's semi-major axis, and $T_{\star}$ and $R_{\star}$ are the effective temperature and radius of the host star.  $A_{eff}$ is equivalent to the Bond albedo given the assumption that the planet's emissivity in the MIRI LRS bandpass is unity.  This assumption is roughly true for regolith surfaces (e.g., \citealt{paragas2025new}) but it is unclear whether it would apply for a reflective atmosphere.  A value of $\mathcal{R}$ near one is indicative of a planet with little-to-no heat redistribution and a low Bond albedo --- e.g., a `bare rock' lacking a thick or reflective atmosphere.  This ratio is converted into a MIRI LRS white light eclipse depth via,
\begin{equation} \label{eq:FpFs}
    \frac{F_{p}}{F_{\star}}=\left(\frac{R_{p}}{R_{\star}}\right)^{2}\times\frac{\int_{\lambda_{min}}^{\lambda_{max}}\frac{\pi B_{\lambda}(\mathcal{R}\times T_{max}(T_{\star},a/R_{\star}))}{hc/\lambda}W_{inst,\lambda}d\lambda}{\int_{\lambda_{min}}^{\lambda_{max}}\frac{M_{\lambda}(T_{\star},\log{(g)},[\mathrm{M}])}{hc/\lambda}W_{inst,\lambda}d\lambda},
\end{equation}
where $B_{\lambda}$ is the blackbody Planck function, [M] is the log$_{10}$ stellar metallicity relative to solar, $\log(g)$ is the stellar surface gravity, $M_{\lambda}$ is the model stellar flux, determined from PHOENIX stellar models \citep{husser13} interpolated using \texttt{pysynphot} \citep{pysynphot}.  We compared the median PHOENIX model using our ExoFASTv2-retrieved stellar parameters to the \texttt{jwst stage3} calibrated stellar spectrum and saw very good agreement in stellar flux to within $\sim2\%$.  This is comparable to the uncertainty expected due to $T_{\star}$ and distance uncertainties and indicates that the PHOENIX model is a good approximation of the host star spectrum. $W_{inst,\lambda}$ is the MIRI LRS throughput (photon to electron conversion efficiency) function, which we determine from \texttt{Pandeia~4.0} \citep{pontoppidan16}, and $\lambda_{min}$ \& $\lambda_{max}$ are the wavelength bounds of our white light curve (5.06$\,\mu$m and 10.55$\,\mu$m, respectively.

\begin{figure}
    \centering
    \includegraphics[width=1.0\linewidth]{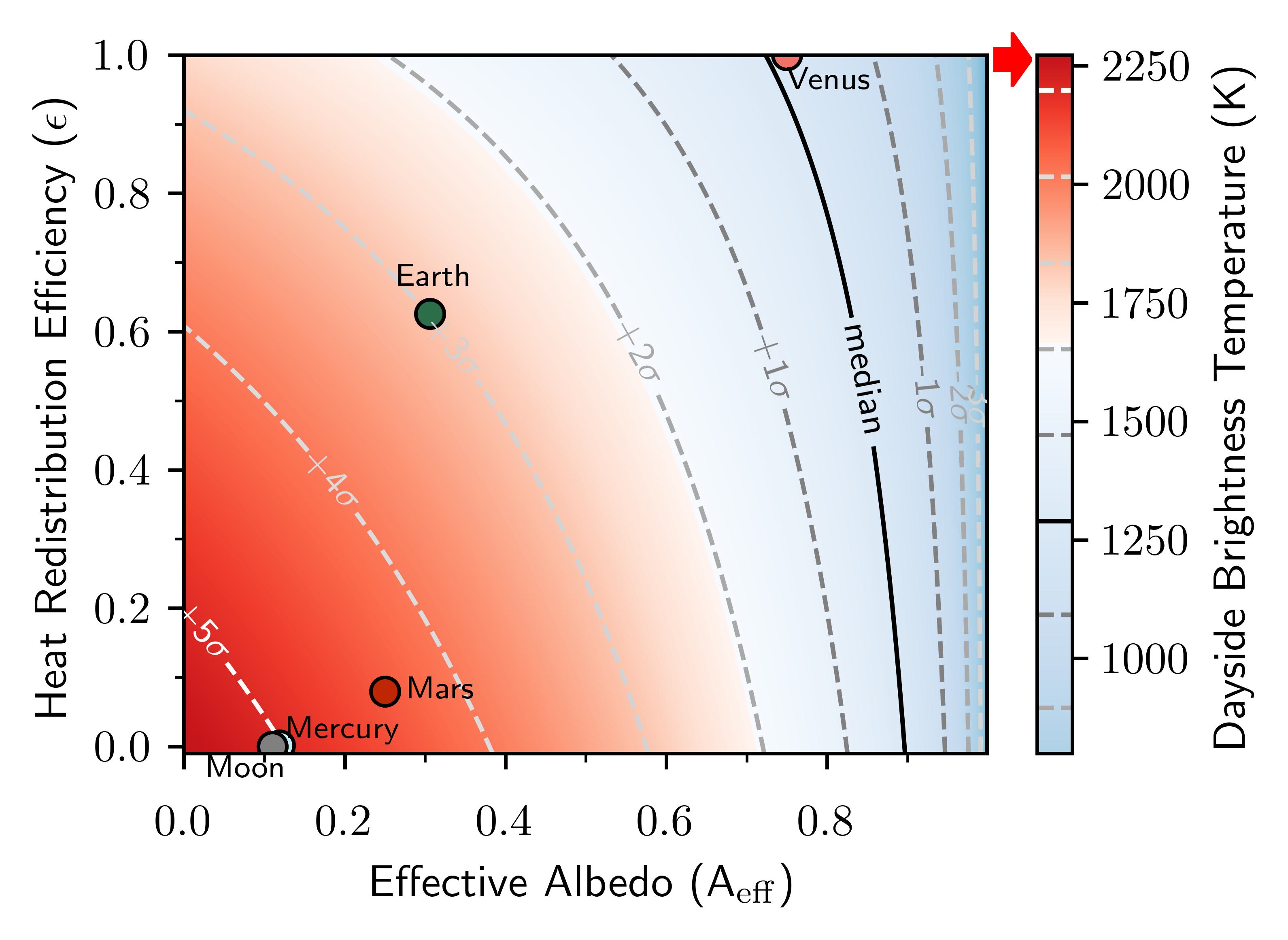}
    \caption{Our retrieved brightness temperature ratio and uncertainties for HD~3167~b in context of Solar System terrestrial bodies.  Estimates for planetary Bond albedo and heat redistribution efficiency are taken from \citet{xue2024jwst}. Dashed lines represent 1-, 2-, 3-, 4-, and 5-$\sigma$ confidence intervals, and the red arrow represents the maximum possible dayside temperature of a perfect blackbody. Our data are inconsistent with a thin Mars-like atmosphere at roughly 4$\,\sigma$ and are most consistent with a thick, highly reflective atmosphere like that of Venus.}
    \label{fig:heatmap}
\end{figure}

Following \citet{xue2024jwst}, we directly use the posterior chain from our global \texttt{ExoFASTv2} fit to derive a posterior distribution for $\mathcal{R}$. This approach better deals with correlated parameters compared to using Gaussian priors for orbital parameters done in previous studies (although the difference in $\mathcal{R}$ is minor, see \citealt{coy2025population} Appendix Table A1). We derive $\mathcal{R}=0.57^{+0.08}_{-0.09}$ (and a corresponding brightness temperature of $1300^{+180}_{-200}$\,K). This low value of $\mathcal{R}$ is 5\,$\sigma$ inconsistent with a dark, bare rock, like the Moon or Mercury.  We show HD~3167~b's brightness temperature ratio in context of estimates for Solar System bodies in Figure \ref{fig:heatmap}, showing that it is most consistent with a thick, highly reflective atmosphere like Venus'.  Notably, efficient heat redistribution alone is not enough to explain our observations; full heat redistribution still requires a high effective albedo of $A_{eff}=0.72^{+0.14}_{-0.19}$, similar to the Bond albedo of Venus.  This could indicate the presence of highly reflective clouds further cooling the dayside.  While future phase curve observations may be able to uniquely constrain heat redistribution and Bond albedo, the very small thermal emission signal size of HD~3167~b may prevent this.

\begin{figure*}
    \centering
    \includegraphics[width=0.75\linewidth]{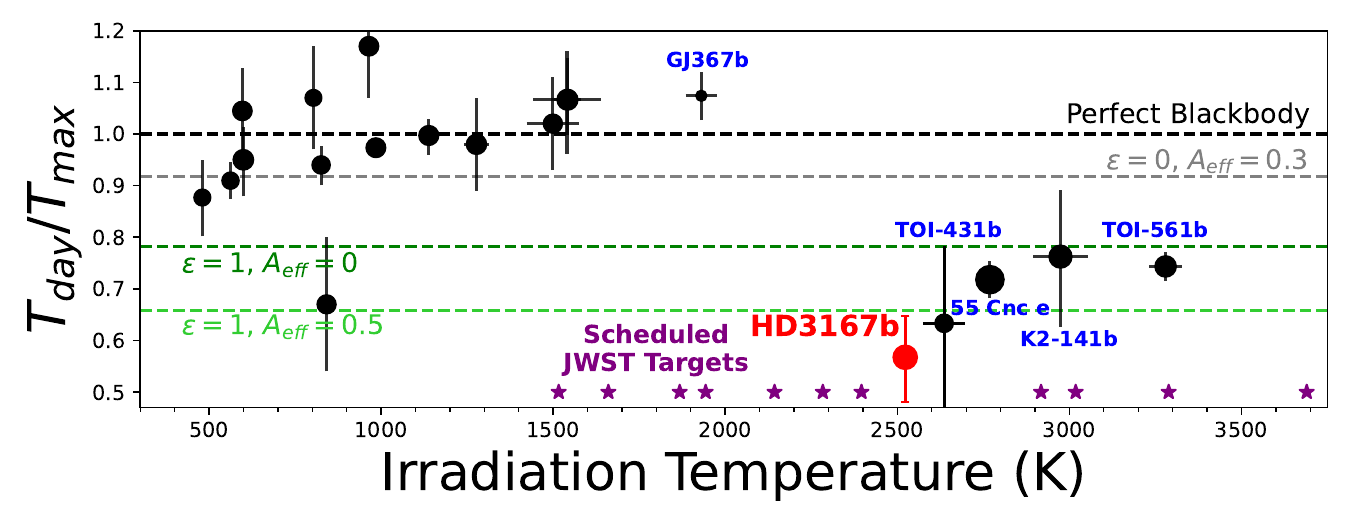}
    \caption{HD~3167~b's measured brightness temperature ratio in context of other Earth-sized ($<1.9R_{\oplus}$) planets observed in thermal emission.  Dashed lines represent different assumptions about the planet's heat redistribution efficiency $\varepsilon$ and effective albedo $A_{eff}$ and point size represents planet surface area.  HD~3167~b better constrains a potential dichotomy between lava worlds with atmospheres and warm rocky M star planets without.  Future emission observations of lava worlds with JWST, shown as purple stars (GJ~9827~b, TOI-1442~b, TOI-1075~b, TOI-1416~b, TOI-500~b, TOI-1807~b, HD~20329~b: JWST GO 4818, PI Weiner Mansfield,  TOI-6255~b, TOI-2431~b, TOI-2260~b: JWST GO 8864: PI Dang, and WASP-47~e: JWST GO 3615, PI Zieba), will better constrain this potential transition temperature. See Appendix Section \ref{ap:tempratio} for a discussion of data used in these plots. 
    }
    \label{fig:trend}
\end{figure*}

We show HD~3167~b's brightness temperature ratio in the broader context of all Earth-sized ($\lesssim1.9R_{\oplus}$) planets observed in thermal emission in Figure \ref{fig:trend}. Among lava worlds ($T_{irr}\gtrsim1500\,$K), HD~3167~b is the coldest with evidence for an atmosphere.  Interestingly, the onset of $\mathcal{R}$ values less than unity, indicating the likely onset of atmospheres, seems to be correlated with temperature, not density: GJ 1252 b ($T_{irr}=1540\,$K), a colder lava world with low uncompressed density relative to Earth, seemingly lacks a thick atmosphere from thermal emission measurements \citep{Crossfield22}, whereas all lava worlds hotter than HD~3167~b thus far show evidence for atmospheres over a range of uncompressed densities.  Future emission observations will be able to confirm or deny this potential transition temperature and whether the onset of atmospheres is correlated with planet bulk density.

\subsection{Atmospheric Forward Modeling and Prospects for Future Characterization}

To model potential atmospheric emission spectra, we use the \texttt{GENESIS} radiative-convective atmospheric model \citep{Gandhi2017_GENESIS,Piette2020_GENESIS}, adapted for lava world atmospheres in \citet{piette2023rocky}. \texttt{GENESIS} is coupled with the \texttt{FastChem} chemical equilibrium code \citep{Stock2022_fastchem} and the \texttt{VapoRock} magma ocean outgassing code \citep{Wolf2022}. The combined model self-consistently solves for the thermochemical equilibrium composition of an atmospheric layer composed of a mixture of volatiles and vaporized rock.  Following \citet{piette2023rocky}, we test volatile mixtures of 100\% H$_2$O and 67\% O$_2$ with 33\% CO$_2$, and assume a melt pool composition of bulk silicate Earth (45.97\% SiO$_2$, 36.66\% MgO, 8.24\% FeO, 4.77\% Al$_2$O$_3$, 3.78\% CaO, 0.35\% Na$_2$O, 0.18\% TiO$_2$ and 0.04\% K$_2$O, \citealt{schaefer2009chemistry}).  See Appendix Section \ref{ap:genesisopac} for a discussion of opacities used in our models.

HD~3167~b's bulk density disfavors a pure rock vapor atmosphere, as the expected equilibrium vapor pressure would not result in an atmosphere thick enough to noticeably affect its radius/density. \citet{curry2025chemical} predict a silicate vapor pressure of only 20\,mbar at a temperature (2600\,K)  similar to HD~3167~b's substellar point ($\sim$2500\,K), dropping to 0.4\,mbar if the magma pool is allowed to fractionally vaporize and evolve over time.  However, these numbers are highly sensitive to various unknowns like the temperature at the base of the atmosphere, the bulk composition of the planet's crust, and the amount of vertical mixing.  We thus test two end-member cases of a relatively low-(1\%) and high-(30\%) rock vapor fraction. The remaining fraction is made up of volatiles; either pure H$_2$O or the O$_2$~+~CO$_2$ mixture described above.  As noted in \citet{piette2023rocky}, the first-order effect of including more rock vapor is that the atmosphere becomes more isothermal, muting absorption features. Note that for simplicity these models do not include condensation effects, such that at higher rock vapor fractions and cooler temperatures the rock vapor may be supersaturated.

\begin{figure*}
    \centering
    \includegraphics[width=0.85\linewidth]{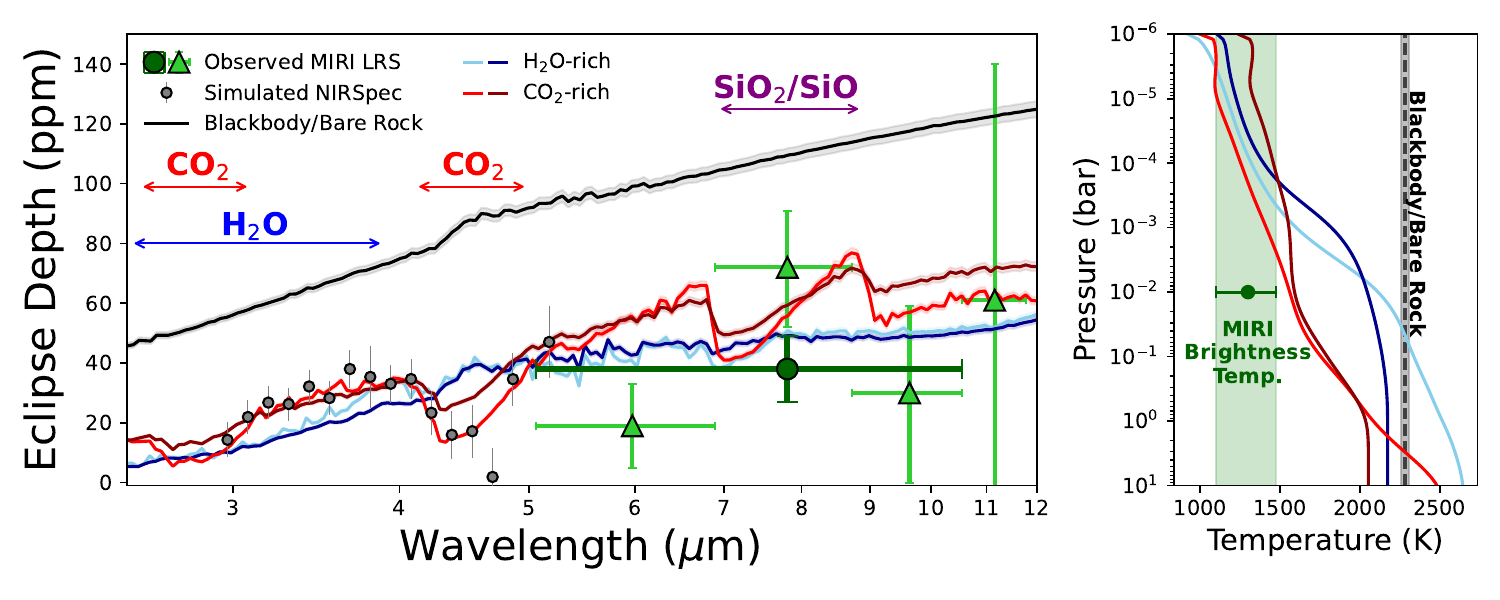}
    \caption{(left) Models of potential emission spectra of HD~3167~b's atmosphere computed by \texttt{GENESIS} \citep{piette2023rocky}, shown alongside our global fit MIRI LRS white light (5.06--10.55$\,\mu$m, excluding the shadowed region) eclipse depth (dark green circle) and \texttt{SPARTA} low-resolution eclipse spectrum (green triangles). Models are binned to a resolving power of $R=100$ for visual clarity. These models assume a volatile component of 67\%\,O$_{2}$+33\%\,CO$_{2}$ (red) and 100\%\,H$_2$O (blue), with dark and light colors indicating high (30\%) and low (1\%) rock vapor fractions, respectively.  Shading represents the uncertainty due to planet-to-star radius ratio uncertainty.  Grey points represent simulated NIRSpec G395H data at $R=15$ assuming 5 total eclipses, showing how future emission observations may be able to distinguish between different volatile-rich scenarios. (right) Temperature-pressure profiles for each atmospheric model, including the maximum theoretical temperature (black) and measured MIRI LRS brightness temperature (green).}
    \label{fig:models}
\end{figure*}

For these models, we assume $\varepsilon=1$ (full heat redistribution to the nightside) and a Bond albedo of $A_{B}=0.3$ for H$_2$O-rich atmospheres and $A_{B}=0.4$ for CO$_2$-rich atmospheres to match our observed flux in MIRI LRS. We fix the surface pressure to 100 bar.  Surface pressures deeper than the photosphere, as is the case here, do not significantly affect the emergent spectrum. These models, alongside our \texttt{SPARTA} eclipse depths, are shown in Figure \ref{fig:models}.

While our single MIRI LRS eclipse is unable to identify potential molecular features, future multi-eclipse observations may be able to.  \citet{piette2023rocky} showed that the NIRSpec G395H wavelength range (2.9-5.2\,$\mu$m) is particularly powerful in distinguishing between volatile-rich (CO$_{2}$/H$_{2}$O) and silicate-rich atmospheres.  We simulate the expected precision of future NIRSpec G395H observations using \texttt{PandExo} \citep{batalha2017pandexo}, assuming a baseline-to-eclipse ratio of 2:1.  We assume 5 total eclipses, corresponding to 24 hours of total observing time, and inflate \texttt{PandExo} errorbars by 20\% based on recent performance compared to \texttt{PandExo} \citep{alderson2024jwst}.  We show in Figure \ref{fig:models} that an H$_2$O-rich atmosphere should be distinguishable from a CO$_2$-rich one with these proposed observations.

We note that our forward models do not consider the effects of reflective silicate clouds, which could increase the planet's Bond albedo while muting potential spectral features.  Highly reflective (likely silicate) clouds with a visible geometric albedo of $A_{g,\rm{CHEOPS}}\sim0.73$ have been observed for LTT~9779~b \citep{saha2025high}, a hot Neptune at a similar irradiation temperature (2800\,K) and thus may be present in HD~3167~b's atmosphere causing its high effective albedo.  Such clouds may be observable via a high geometric albedo in the visible. \citet{vanderburg2016two} were able to place an upper limit on HD~3167~b's Kepler secondary eclipse depth of 12\,ppm, corresponding to an upper limit on the Kepler-bandpass geometric albedo of $A_{g}<0.8$ [$F_{p,ref}/F_{\star}=A_{g}(R_{p}/a)^{2}=15\mathrm{\,ppm}\times A_{g}$], as the thermal contribution in the Kepler bandpass should be negligible.  This upper limit is unfortunately not very informative of the cause of HD~3167~b's high effective albedo. Future observations of HD~3167~b's secondary eclipse with CHEOPS or TESS may be able to better constrain the planet's geometric albedo, although the expected signal size is smaller than the precision achieved with CHEOPS in \citet{bourrier2022cheops}.

\subsection{Surface False Positives}
Our low dayside brightness temperature may alternatively be explained by no/little heat redistribution and a high-albedo surface.  This albedo, however, would have to be extremely high at $A_{eff}=0.90^{+0.05}_{-0.07}$. Although high-quality reflectance spectra of molten silicates at temperatures relevant to HD~3167~b's surface do not yet exist, experiments have shown that molten metal oxides exhibit much lower reflectance than their solid counterparts \citep{dvurechensky1979spectral,petrov2007}. Reflectance spectra of quenched lava glasses of varying composition have also suggested low ($\lesssim0.1$) albedo for lava world surfaces \citep{essack2020low}. This is likely due to regolith particles losing their scattering properties as they liquefy, as solid slabs also show much lower reflectivity than fine-grained regolith \citep{paragas2025new}.

Contrarily, several studies have suggested that the \textit{emissivity} ($\varepsilon$), a proxy for the reflectivity using Kirchoff's law ($\varepsilon_{\lambda}\approx1-R_{\lambda}$, where $R$ is the wavelength-dependent hemispherical reflectance), decreases significantly with increasing temperature, implying that silicates may become more reflective at high temperatures \citep{thompson2021quantitative,fortin2024lava}.  However, these measurements are limited to the mid-infrared ($>2.5\,\mu$m) where there is little incoming starlight, and converting these data to reflectance requires assuming that molten silicates are opaque, which is likely not true for silicate melt samples in the near infrared (e.g., \citealt{ferkl2026heat}). Precise measurements of the \underline{direct} reflectance spectra of molten lava will be required to determine whether the molten surfaces of lava worlds can be shiny enough to significantly cool their daysides. However, current evidence points towards low albedo surfaces for lava worlds, incompatible with our observations.

\section{Conclusions} \label{sec:conc}

We present observations of HD~3167~b, a super-Earth lava world, that indicate the presence of a reflective and/or thick atmosphere.  HD~3167~b joins a growing family of ultra hot lava worlds (TOI-431 b, K2-141 b, 55 Cancri e, and TOI-561 b) that show evidence for atmospheres from thermal emission observations. Notably, HD~3167~b is the least irradiated of these planets, better constraining a potential `transition temperature' of unknown cause where thick or highly reflective atmospheres on lava worlds become common to $2000\,\mathrm{K}\lesssim T_{irr,\rm{transition}}\lesssim2500\,$K.  Future emission observations will be required to constrain the composition of its atmosphere.

\begin{acknowledgments} 

This work is based on observations with the NASA/ESA/CSA JWST. The data were obtained from the Mikulski Archive for Space Telescopes at the Space Telescope Science Institute (STScI), which is operated by the Association of Universities for Research in Astronomy, Incorporated, under NASA contract NAS5-03127. These observations are associated with program  JWST GO 4818. The specific observations analyzed can be accessed via \dataset[doi:10.17909/zdcn-4q44]{https://doi.org/10.17909/zdcn-4q44}. Support for this program was provided through a grant from STScI. This research has made use of the NASA Exoplanet Archive, which is operated by the California Institute of Technology, under contract with the National Aeronautics and Space Administration under the Exoplanet Exploration Program. A.A.A.P. acknowledges funding from a UK Science and Technology Facilities Council (STFC) Small Award, grant number UKRI/ST/B001171/1.  M.Z. is funded by the Heising-Simons Foundation's 51 Pegasi b fellowship. D.D.B.K. is supported by NSFC grant 12473064. R.L. is funded by the European Union (ERC, THIRSTEE, 101164189) and acknowledges financial support from the Severo Ochoa grant CEX2021-001131-S funded by MCIN/AEI/10.13039/501100011033. Views and opinions expressed are however those of the authors only and do not necessarily reflect those of the European Union or the European Research Council. Neither the European Union nor the granting authority can be held responsible for them. 

The authors thank Huan-Yu Teng for sharing their N-body simulations of the HD~3167 system. We thank Michael Radica for useful discussions regarding the interpretation of the PCA components in our MIRI LRS data reduction. We thank an anonymous reviewer for helpful suggestions to improve the manuscript.

\end{acknowledgments}




%
\facilities{JWST (MIRI LRS), TESS, HST (WFC3), Kepler, Spitzer, Keck/HIRES, APF/Levy, ESO3.6m/HARPS, TNG/HARPS-N}

\software{\texttt{numpy} \citep{van2011numpy}, \texttt{scipy} \citep{2020SciPy-NMeth}, \texttt{matplotlib} \citep{Hunter:2007}, \texttt{dynesty} \citep{speagle2020dynesty}, \texttt{Eureka!} \citep{bell_eureka_2022}, \texttt{SPARTA} \citep{kempton2023reflective}, \texttt{ExoTEDRF} \citep{radica2024exoTEDRF}, \texttt{ExoFASTv2} \citep{Eastman:2019}}



\appendix

\twocolumngrid

\section{Sensitivity to Systematics Model}
\label{ap:sys}

Due to the small anticipated signal size of HD~3167~b's eclipse ($\lesssim$\,110 ppm), the choice of systematics model can affect the retrieved eclipse depth.  The systematics of MIRI LRS are typically modeled as a combination of an exponential ramp, a linear slope, and/or a quadratic component. Further decorrelation with the position and width of the spectral trace is also often required for small signals.  The first $\sim$30-60 minutes of MIRI observations are often trimmed as these simple systematics models fail to accurately capture the beginning of the observations where the ramp is typically most severe.  In reality, the systematics are a complex summation of thousands of individual pixel-level ramps, each of which can have 5 or more unique exponential components \citep{dyrek2024transiting}.

We tested trimming from 600 (16 minutes) up to 1800 (48 minutes) integrations from the white light curve in intervals of 200.  Our observation shows a very steep ramp in the first $\sim10$ minutes of the observations that is not well-captured by a single ramp or quadratic model.  For each number of trimmed integrations, we test a linear-only, linear plus exponential, and quadratic ramp model. We compared the normalized log-likelihood $\ln\mathcal{L}/(n - k)$, where $n$ and $k$ are the number of fit data points and degrees of freedom, respectively, and the Bayesian Information Criterion/Corrected Akaike Information Criterion (BIC/AICc) for each best-fit model.  We determined that trimming 1200 integrations (32 minutes) gives the optimal balance of goodness-of-fit and information criteria.  Both eccentric and circular models show eclipse depths 1\,$\sigma$ consistent with each other when trimming 1000 or more integrations, with the eccentric fits giving slightly deeper eclipses. A linear-only model finds much deeper eclipse depths ($\sim110$\,ppm) but shows clear ramp-like residuals and has the worst goodness of fit. The quadratic model gives the best BIC and AICc and seems to more accurately capture the systematic behavior than the exponential ramp+linear model for each pipeline, which fits ramp timescales much longer ($\gtrsim18$ hours) than the observation itself. The ramp+linear model finds eclipse depths consistent with our adopted quadratic model ($40^{+12}_{-11}$ ppm for \texttt{SPARTA}).

\section{Allowing Free Eccentricity}\label{ap:ecc}

The orbits of ultra short period planets are generally thought to be very near circular due to their extremely short tidal circularization timescales. General relativity further circularizes these orbits (e.g., \citealt{marzari2020secular}).  However, it is possible that HD~3167~b migrated to its current position very recently.  Therefore, a non-zero eccentricity is conceivable.

We tested including $e\cos\omega$ and $e\sin\omega$ as free parameters in our \texttt{SPARTA} light curve fitting with Gaussian priors taken from \citet{bourrier2022cheops} informed by RV and high-precision transit photometry ($e\cos\omega=-0.015\pm0.023$, $e\sin\omega=0.030\pm0.034$).
We retrieved posteriors of $e\cos\omega=0.0045^{+0.0039}_{-0.0017}$ and $e\sin\omega=0.079^{+0.025}_{-0.039}$, both more than $2\sigma$ inconsistent with zero.  This is driven by the longer-than expected eclipse duration ($1.786^{+0.044}_{-0.068}$ hr compared to a transit duration of $1.609^{+0.017}_{-0.014}$ hr), and the slight offset of mid-eclipse relative to a circular orbit ($4.5^{+3.5}_{-1.4}$ min compared to the 1.5 min uncertainty on ephemeris).  The preference for an extended eclipse duration can be clearly seen over all reduction pipelines from the post-eclipse binned data point in Figure \ref{fig:lightcurves}.

However, this behavior is not consistent across different pipelines and/or assumptions, and the corresponding eccentricity derived from \texttt{SPARTA} ($e=0.079^{+0.025}_{-0.036}$) does not surpass the 2.45\,$\sigma$ threshold identified by \citet{lucy1971spectroscopic} to be considered `inconsistent with zero'.  For example, \texttt{Eureka!} finds values ($e\sin\omega=0.050^{+0.034}_{-0.032}$, $e\cos\omega=0.0032^{+0.0117}_{-0.0076}$) more consistent with a circular orbit, and when allowing for a free eccentricity \texttt{ExoFASTv2} finds multiple modes for the mid-eclipse timing, some with negative eclipse depths, with the most favored being near the expected circular timing.  In addition, the relatively large values of $e\sin\omega$ compared to $e\cos\omega$ imply an argument of periastron very near 90$\degree$ ($\omega=86.8^{+1.5}_{-4.2}\,\degree$), which is very unlikely assuming a random distribution for $\omega$.

Given that our eclipse signal is small and only detected at $\lesssim3.5\sigma$, we believe that our retrieved non-zero eccentricity, mainly driven by the longer-than-expected eclipse duration, is caused by a slight deviation in the behavior of the systematic ramp near the end of the eclipse.  These eccentric fits result in eclipse depths within 1$\,\sigma$ of our fiducial circular fits ($48\pm13$ ppm for \texttt{SPARTA}, $37^{+16}_{-17}$ ppm for \texttt{Eureka!}, $34\pm15$ ppm for \texttt{exoTEDRF}), reinforcing our inference of a shallow eclipse.

\section{Global Fit Results}\label{ap:exofast}
Full results of our global fit using \texttt{ExoFASTv2} described in Section \ref{sec:global} are shown in Table \ref{tab:HD3167.circ.}. We are able to use the precise stellar density derived from our JWST eclipse to improve upon the systematic floors in stellar parameter precision identified by \citet{tayar2022}. Following the procedure from \citet{mahajan2024using}, we achieve a precision of 2.3\% in $R_{\star}$\ and 1.3\% in $T_{eff}$, compared to the 4.2\% and 2.4\% systematic floors suggested by \citet{tayar2022}. These are nominally less precise than the uncertainties presented by \citet{bourrier2022cheops} (0.7\% and 1.4\%, respectively). However, their values seem to ignore the systematic floors identified by \citet{tayar2022} in the infrared flux method used to derive their stellar parameters. Regardless, our derived stellar parameters ($R_{\star}=0.865^{+0.022}_{-0.017}\,R_{\odot}$, $T_{eff}=5350^{+66}_{-71}$\,K, $M_{\star}=0.864^{+0.039}_{-0.034}\,M_{\odot}$) are highly consistent with those presented in \citet{bourrier2022cheops} ($R_{\star}=0.865\pm0.006\,R_{\odot}$, $T_{eff}=5300\pm73$\,K, $M_{\star}=0.852^{+0.026}_{-0.015}\,M_{\odot}$).  Notably, we retrieve a planet density for HD~3167~b ($6.49^{+0.52}_{-0.60}$\,g\,cm$^{-3}$) 2$\,\sigma$ higher than that reported in \citet{bourrier2022cheops} ($5.5\pm0.5$\,g\,cm$^{-3}$), closer to the density expected of an Earth-like composition.  Our global fit revises the planet's uncompressed density compared to Earth (based on \citealt{zeng2019}, see Appendix Section \ref{ap:tempratio}) from $\rho=0.67^{+0.07}_{-0.11}\,\rho_{\oplus}$ to $\rho=0.82\pm0.05\,\rho_{\oplus}$.  This is due to a combination of our lower retrieved radius ($R=1.601^{+0.048}_{-0.036}\,R_{\oplus}$) compared to \citet{bourrier2022cheops} ($R=1.627^{+0.083}_{-0.058}\,R_{\oplus}$), who fit for separate radii over different instrument bandpasses, and our slightly higher retrieved mass ($M=4.84^{+0.27}_{-0.24}\,M_{\oplus}$) compared to $M=4.73^{+0.28}_{-0.29}\,M_{\oplus}$ from \citet{bourrier2022cheops} due to the addition of new RV data from \citet{bonomo2023cold} (who report $M=4.97^{+0.24}_{-0.23}\,M_{\oplus})$.

\providecommand{\bjdtdb}{\ensuremath{\rm {BJD_{TDB}}}}
\providecommand{\tjdtdb}{\ensuremath{\rm {TJD_{TDB}}}}
\providecommand{\feh}{\ensuremath{\left[{\rm Fe}/{\rm H}\right]}}
\providecommand{\teff}{\ensuremath{T_{\rm eff}}}
\providecommand{\teq}{\ensuremath{T_{\rm eq}}}
\providecommand{\ecosw}{\ensuremath{e\cos{\omega_*}}}
\providecommand{\esinw}{\ensuremath{e\sin{\omega_*}}}
\providecommand{\msun}{\ensuremath{\,M_\Sun}}
\providecommand{\rsun}{\ensuremath{\,R_\Sun}}
\providecommand{\lsun}{\ensuremath{\,L_\Sun}}
\providecommand{\mj}{\ensuremath{\,M_{\rm J}}}
\providecommand{\rj}{\ensuremath{\,R_{\rm J}}}
\providecommand{\me}{\ensuremath{\,M_{\rm E}}}
\providecommand{\re}{\ensuremath{\,R_{\rm E}}}
\providecommand{\fave}{\langle F \rangle}
\providecommand{\fluxcgs}{10$^9$ erg s$^{-1}$ cm$^{-2}$}

\begin{longrotatetable}
\small
\startlongtable
\begin{deluxetable*}{lccccc}
\tablenum{A1}
\tabletypesize{\scriptsize}
\tablecaption{Median values and 68\% confidence interval for our global analysis of the HD~3167 system assuming $e_{b}=0$, created using \texttt{EXOFASTv2}}
\tablehead{\colhead{~~~Parameter} & \colhead{Description} & \multicolumn{4}{c}{Values}}
\startdata
\smallskip\\\multicolumn{2}{l}{Stellar Parameters:}&HD~3167\smallskip\\
~~~~$M_*$\dotfill &Mass (\msun)\dotfill &$0.864^{+0.039}_{-0.034}$\\
~~~~$R_*$\dotfill &Radius (\rsun)\dotfill &$0.865^{+0.022}_{-0.017}$\\
~~~~$R_{*,SED}$\dotfill &Radius$^{1}$ (\rsun)\dotfill &$0.871\pm0.015$\\
~~~~$L_*$\dotfill &Luminosity (\lsun)\dotfill &$0.553\pm0.020$\\
~~~~$\rho_*$\dotfill &Density (cgs)\dotfill &$1.903^{+0.088}_{-0.140}$\\
~~~~$\log{g}$\dotfill &Surface gravity (cgs)\dotfill &$4.504^{+0.016}_{-0.024}$\\
~~~~$T_{\rm eff}$\dotfill &Effective temperature (K)\dotfill &$5350^{+66}_{-71}$\\
~~~~$T_{\rm eff,SED}$\dotfill &Effective temperature$^{1}$ (K)\dotfill &$5338^{+58}_{-60}$\\
~~~~$[{\rm Fe/H}]$\dotfill &Metallicity (dex)\dotfill &$0.037^{+0.078}_{-0.081}$\\
~~~~$[{\rm Fe/H}]_{0}$\dotfill &Initial Metallicity$^{2}$ \dotfill &$0.057^{+0.076}_{-0.078}$\\
~~~~$Age$\dotfill &Age (Gyr)\dotfill &$8.4^{+3.3}_{-3.5}$\\
~~~~$EEP$\dotfill &Equal Evolutionary Phase$^{3}$ \dotfill &$360^{+22}_{-18}$\\
~~~~$A_V$\dotfill &V-band extinction (mag)\dotfill &$0.050^{+0.030}_{-0.033}$\\
~~~~$\sigma_{SED}$\dotfill &SED photometry error scaling \dotfill &$2.18^{+0.78}_{-0.48}$\\
~~~~$\varpi$\dotfill &Parallax (mas)\dotfill &$21.171\pm0.021$\\
~~~~$d$\dotfill &Distance (pc)\dotfill &$47.234^{+0.047}_{-0.048}$\\
\smallskip\\\multicolumn{2}{l}{Planetary Parameters:}&HD~3167~b&HD~3167~c&HD~3167~d&HD~3167~e\smallskip\\
~~~~$P$\dotfill &Period (days)\dotfill &$0.95965451^{+0.00000022}_{-0.00000021}$&$29.846511\pm0.000015$&$8.3967^{+0.0024}_{-0.0025}$&$91.90^{+0.30}_{-0.29}$\\
~~~~$R_P$\dotfill &Radius (\re)\dotfill &$1.601^{+0.048}_{-0.036}$&$2.983^{+0.083}_{-0.068}$& -- & -- \\
~~~~$M_P$\dotfill &Mass (\me)\dotfill &$4.844^{+0.270}_{-0.254}$&$11.41^{+0.858}_{-0.826}$&$7.25^{+9.85}_{-2.45}$&$14.94^{+15.89}_{-4.770}$\\
~~~~$T_C$\dotfill &Observed Time of conjunction$^{4}$ (\bjdtdb)\dotfill &$2457394.37378^{+0.00041}_{-0.00043}$&$2457394.97718\pm0.00056$&$2457745.09^{+0.19}_{-0.18}$&$2457735.9^{+2.8}_{-3.4}$\\
~~~~$T_C$\dotfill &Model Time of conjunction$^{4,5}$ (\tjdtdb)\dotfill &$2457394.37367^{+0.00041}_{-0.00043}$&$2457394.97611\pm0.00056$&$2457745.09^{+0.19}_{-0.18}$&$2457735.9^{+2.8}_{-3.4}$\\
~~~~$T_T$\dotfill &Model time of min proj sep$^{5,6,7}$ (\tjdtdb)\dotfill &$2458925.98226^{+0.00026}_{-0.00025}$&$2458499.29702^{+0.00013}_{-0.00014}$&--&--\\
~~~~$T_0$\dotfill &Obs time of min proj sep$^{6,8,9}$ (\bjdtdb)\dotfill &$2458925.98237^{+0.00026}_{-0.00025}$&$2458499.29808\pm0.00013$& -- & -- \\
~~~~$a$\dotfill &Semi-major axis (AU)\dotfill &$0.01814^{+0.00027}_{-0.00024}$&$0.1794^{+0.0027}_{-0.0024}$&$0.0770^{+0.0011}_{-0.0010}$&$0.3797^{+0.0057}_{-0.0051}$\\
~~~~$i$\dotfill &Inclination (Degrees)\dotfill &$87.8^{+1.5}_{-1.6}$&$89.433^{+0.094}_{-0.077}$&$39^{+34}_{-24}$&$41^{+32}_{-23}$\\
~~~~$e$\dotfill &Eccentricity \dotfill & 0 (fixed) &$0.040^{+0.044}_{-0.028}$&$0.35^{+0.13}_{-0.12}$&$0.128^{+0.110}_{-0.087}$\\
~~~~$\omega_*$\dotfill &Arg of periastron (Degrees)\dotfill & 90 (fixed) &$-69^{+71}_{-49}$&$62^{+19}_{-15}$&$0^{+73}_{-60}$\\
~~~~$\dot{\omega}_{\rm GR}$\dotfill &Computed GR precession ($^\circ$/century)\dotfill &$19.33^{+0.58}_{-0.51}$&$0.0631^{+0.0019}_{-0.0017}$&$0.595^{+0.082}_{-0.047}$&$0.00988^{+0.00049}_{-0.00035}$\\
~~~~$T_{\rm eq}$\dotfill &Equilibrium temp$^{10}$ (K)\dotfill &$1781^{+18}_{-19}$&$566.6^{+5.8}_{-5.9}$&$864.7^{+8.9}_{-9.1}$&$389.4^{+4.0}_{-4.1}$\\
~~~~$K$\dotfill &RV semi-amplitude (m/s)\dotfill &$3.46\pm0.16$&$2.59^{+0.18}_{-0.17}$&$1.70\pm0.18$&$1.55\pm0.17$\\
~~~~$R_P/R_*$\dotfill &Radius of planet in stellar radii \dotfill &$0.01698^{+0.00019}_{-0.00018}$&$0.03162^{+0.00024}_{-0.00021}$& -- & -- \\
~~~~$a/R_*$\dotfill &Semi-major axis in stellar radii \dotfill &$4.525^{+0.069}_{-0.110}$&$44.75^{+0.68}_{-1.10}$&$19.21^{+0.29}_{-0.49}$&$94.7^{+1.5}_{-2.4}$\\
~~~~$\delta$\dotfill &$\left(R_P/R_*\right)^2$ (ppm) \dotfill &$288.4^{+6.4}_{-6.0}$&$1000^{+15}_{-13}$& -- & -- \\
~~~~$\tau$\dotfill &In/egress transit duration (hours)\dotfill &$0.0286^{+0.0018}_{-0.0008}$&$0.186^{+0.022}_{-0.015}$& -- & --\\
~~~~$T_{14}$\dotfill &Total transit duration (hours)\dotfill &$1.636^{+0.014}_{-0.015}$&$4.878^{+0.023}_{-0.019}$& -- & --\\
~~~~$b$\dotfill &Transit impact parameter \dotfill &$0.17^{+0.12}_{-0.11}$&$0.452^{+0.078}_{-0.080}$& -- & -- \\
~~~~$\rho_P$\dotfill &Density (cgs)\dotfill &$6.49^{+0.52}_{-0.60}$&$2.36^{+0.22}_{-0.23}$& -- & -- \\
~~~~$logg_P$\dotfill &Surface gravity (cgs)\dotfill &$3.267^{+0.028}_{-0.032}$&$3.097^{+0.034}_{-0.037}$& -- & -- \\
~~~~$T_S$\dotfill &Observed Time of eclipse$^{4}$ (\bjdtdb)\dotfill &$2457394.85340^{+0.00040}_{-0.00043}$&$2457380.24^{+0.51}_{-0.35}$& -- & -- \\
~~~~$T_S$\dotfill &Model Time of eclipse$^{4,5}$ (\tjdtdb)\dotfill &$2457394.85350^{+0.00041}_{-0.00043}$&$2457380.24^{+0.51}_{-0.35}$& -- & -- \\
~~~~$T_D$\dotfill &Time of desc node (\tjdtdb)\dotfill &$2457394.61359^{+0.00041}_{-0.00043}$&$2457402.78^{+0.61}_{-0.37}$&$2457746.61^{+0.36}_{-0.35}$&$2457669.1^{+4.3}_{-4.2}$\\
~~~~$e\cos{\omega_*}$\dotfill & \dotfill & 0 (fixed) &$0.010^{+0.027}_{-0.018}$&$0.15^{+0.12}_{-0.11}$&$0.071^{+0.110}_{-0.076}$\\
~~~~$e\sin{\omega_*}$\dotfill & \dotfill & 0 (fixed) &$-0.023^{+0.029}_{-0.053}$&$0.31^{+0.10}_{-0.12}$&$0.001^{+0.100}_{-0.099}$\\
~~~~$M_P\sin i$\dotfill &Minimum mass (\me)\dotfill &$4.840^{+0.270}_{-0.254}$&$11.41^{+0.858}_{-0.826}$&$4.545\pm0.477$&$9.789^{+1.112}_{-1.081}$\\
~~~~$d/R_*$\dotfill &Separation at mid transit \dotfill &$4.525^{+0.069}_{-0.110}$&$45.7^{+2.5}_{-1.8}$& -- & -- \\
\smallskip\\\multicolumn{2}{l}{Wavelength Parameters:}&CHEOPS&HST&Kepler&MIRI\smallskip\\
~~~~$u_{1}$\dotfill &Linear limb-darkening coeff \dotfill &$0.33^{+0.22}_{-0.20}$&$0.158^{+0.078}_{-0.082}$&$0.48^{+0.19}_{-0.22}$& -- \\
& & Spitzer &TESS\smallskip\\
& & $0.14^{+0.16}_{-0.10}$&$0.50^{+0.29}_{-0.30}$\\
~~~~$u_{2}$\dotfill &Quadratic limb-darkening coeff \dotfill &$0.56^{+0.23}_{-0.28}$&$0.38\pm0.11$&$-0.02^{+0.30}_{-0.22}$& -- \\
& & $0.01^{+0.12}_{-0.11}$&$0.11^{+0.41}_{-0.36}$\\
~~~~$\delta_{S}$\dotfill &Measured HD~3167~b eclipse depth (ppm)\dotfill &--&--&--&$38\pm11$\\
\smallskip\\\multicolumn{2}{l}{Telescope Parameters:}&APF&HARPS-N&HARPS&HIRES\smallskip\\
~~~~$\gamma_{\rm rel}$\dotfill &Relative RV Offset (m/s)\dotfill &$-0.21\pm0.37$&$0.16\pm0.15$&$-0.18\pm0.32$&$0.51^{+0.42}_{-0.43}$\\
~~~~$\sigma_J$\dotfill &RV Jitter (m/s)\dotfill &$3.38^{+0.30}_{-0.27}$&$1.403^{+0.110}_{-0.097}$&$1.49^{+0.28}_{-0.24}$&$2.99^{+0.36}_{-0.31}$\\
~~~~$\sigma_J^2$\dotfill &RV Jitter Variance \dotfill &$11.4^{+2.1}_{-1.8}$&$1.97^{+0.31}_{-0.26}$&$2.22^{+0.91}_{-0.65}$&$8.9^{+2.3}_{-1.8}$\\
\enddata
\tablecomments{See Table 3 in \citet{Eastman:2019} for a detailed description of all parameters. $^{1}$This value ignores the systematic error and is for reference only. $^{2}$ The metallicity of the star at birth. $^{3}$ Corresponds to static points in a star's evolutionary history. See \S2 in \citet{Dotter:2016}. $^{4}$ Time of conjunction is commonly reported as the ``transit time''. $^{5}$ \tjdtdb is the target's barycentric frame and corrects for light travel time. $^{6}$ Time of minimum projected separation is a more correct ``transit time''. $^{7}$ Use this to model TTVs. $^{8}$ At the epoch that minimizes the covariance between $T_C$ and Period. $^{9}$ Use this to predict future transit times. $^{10}$ Assumes zero albedo and perfect heat redistribution}
\label{tab:HD3167.circ.}
\end{deluxetable*}
\end{longrotatetable}

\section{\texttt{GENESIS} Opacity Sources} \label{ap:genesisopac}

The \texttt{GENESIS} models include UV to infrared opacity from key molecular, atomic and ionic species in the atmosphere. For the molecular species, we calculate absorption cross sections as in \citet{Gandhi2017_GENESIS} using line list data from the ExoMol, HITEMP and HITRAN databases. We consider opacity from the following molecular species and corresponding line lists: H$_2$O \citep{Rothman2010}, CO$_2$ \citep{Rothman2010}, CO \citep{Rothman2010}, SiO \citep{Yurchenko2021}, SiO$_2$ \citep{Owens2020}, AlO \citep{Patrascu2015}, MgO \citep{Li2019}, NaO \citep{Mitev2022}, TiO \citep{McKemmish2019}, O$_2$ \citep{Gordon2017}, OH \citep{Rothman2010}, FeH \citep{Dulick2003,Bernath2020}, NaH \citep{Rivlin2015}, NaOH \citep{Owens2021} and KOH \citep{Owens2021}. For the atomic and ionic species, we use opacities from the \texttt{DACE}\footnote{https://dace.unige.ch/} database, calculated using \texttt{helios-k} \citep{Grimm2021} and data from the Kurucz\footnote{http://kurucz.harvard.edu/} database \citep{Kurucz2018}. We consider opacity from the following atomic and ionic species: Al, Ca, Fe, H, K, Mg, Na, O, Si, Ti, Ca$^+$, and Na$^+$.

\section{Determining Temperature Ratios and Uncompressed Densities}\label{ap:tempratio}

Brightness temperature ratio data used in Figure \ref{fig:trend} are taken from \citet{coy2025population} using PHOENIX stellar models for uniformity. New values for $\mathcal{R}$ are taken directly from the reported values for LHS~1140~c (\citealt{fortune2025hot}, $\mathcal{R}=1.045\pm0.084$), TOI-1468~b (\citealt{valdes2025hot}, $\mathcal{R}=1.17\pm0.10$), GJ~3929~b (\citealt{xue2025jwst}, $\mathcal{R}=1.07\pm0.10$), LTT~3780~b (\citealt{allen2025hot}, $\mathcal{R}=0.98\pm0.09$), TOI-431~b (\citealt{monaghan2025low}, $\mathcal{R}=0.63^{+0.16}_{-0.15}$), 55 Cancri e (MIRI LRS only, \citealt{hu2024secondary}, $\mathcal{R}=0.718\pm0.036$), K2-141~b (\citealt{zieba2022k2}, $\mathcal{R}=0.76\pm0.13$), and TOI-561~b (`\texttt{ExoTiC JEDI 2}' reduction, \citealt{teske2025thick}, $\mathcal{R}=0.727\pm0.028$). Equilibrium/irradiation temperature data are taken from the NASA Exoplanet Archive (accessed March 30, 2026).

We investigated a similar trend of atmosphere absence/presence on lava worlds compared to their uncompressed densities relative to the Earth-like (32.5 wt.\% Fe/Ni-metal and 67.5 wt.\%
MgSiO$_3$-rock) mass-radius models of \citet{zeng2019}.  However, we found that several potentially low-density lava worlds have conflicting mass and/or radius measurements that makes determination of the true density difficult; for GJ~1252~b, masses of $2.09\pm0.56\,M_{\oplus}$ \citep{shporer2020gj}  or $1.32\pm0.28\,M_{\oplus}$ \citep{luque22} lead to uncompressed densities of $1.10^{+0.43}_{-0.34}\,\rho_{\oplus}$ or $0.69^{+0.24}_{-0.19}\,\rho_{\oplus}$, respectively.  We ultimately determined that the current density and thermal emission data are not precise enough to identify statistical trends in atmosphere presence/absence.

\bibliography{main}{}
\bibliographystyle{aasjournalv7}



\end{document}